\documentclass[letterpaper, 10 pt, conference]{ieeeconf}  

\usepackage{amssymb,enumerate}
\usepackage{mathtools}
\usepackage{amsmath}
\usepackage{verbatim}
\usepackage{soul}
\usepackage{ifthen}
\usepackage{xkeyval}
\usepackage{xcolor}
\usepackage{tikz}
\usepackage{calc}
\usepackage{graphicx,psfrag}
\usepackage{fancyheadings}
\usepackage{booktabs}
\usepackage{multirow}
\usepackage{cite}
\usepackage[normalem]{ulem}
\IEEEoverridecommandlockouts                              

\newtheorem{theorem}{Theorem}

\newtheorem{lemma}{Lemma}
\newtheorem{problem}{Problem}

\begin{document}

\title{Cooperative Robust Estimation with Local Performance Guarantees%
\thanks{This work was supported by the Australian Research
  Council under Discovery Projects funding scheme (project
  DP120102152). This work was done while the first author was with the School of
  Engineering and IT, UNSW Canberra, Canberra, Australia.}}

\author{M. Zamani \and V. Ugrinovskii
\thanks{M. Zamani is with the DST Group Australia and V. Ugrinovskii is with School of Engineering and IT, 
        UNSW Canberra, Canberra, Australia, Email: 
        {\tt\small \{m.zammani,v.ugrinovskii\}@gmail.com}}}

\maketitle
\thispagestyle{empty}
\pagestyle{empty}

\begin{abstract}
The paper considers the problem of cooperative estimation for a linear
uncertain plant observed by a network of communicating sensors. We take a
novel approach by treating the filtering problem from the view point of
local sensors while the network interconnections are accounted for via an
uncertain signals modelling of estimation performance of other
nodes. That is, the information communicated between the nodes is
treated as the true plant information subject to perturbations, and each
node is endowed with certain believes about these perturbations during the
filter design. The proposed distributed filter achieves a
suboptimal $H_\infty$ consensus performance. Furthermore,
local performance of each estimator is also assessed given additional
constraints on the performance of the other nodes. These 
conditions are shown to be useful in tuning the desired
estimation performance of the sensor network.
\end{abstract}

\section{Introduction}
The research on cooperative filtering and estimation of networked systems
has gained much momentum during the past decade, aiming at developing
efficient estimation algorithms for large assemblies of networked
sensors~\cite{Olfati2005,SS-2009,NF-2009,U6,LaU1}.   
%
%
The mentioned references reflect the common trend in the 
literature
, where the main
objective is to accomplish a globally optimal or suboptimal estimation
performance of the network. Usually, the performance of individual sensors is 
not considered in such problems. This observation motivates the
question about a relationship between the estimation performance
of the individual filters within a distributed
estimation network and the performance of the overall network. 
This paper considers this problem within the specific framework of distributed
$H_\infty$ consensus estimation~\cite{U6,LaU1}.

Our approach also targets the global convergence problem however not with a
brute-force decoupling of the global solution. 
Rather we define a local objective function 
in terms of uncertain signals capturing the performance of
the other nodes. 
This leads to decoupling of the distributed filter while implicitly
maintaining a meaningful connection to the network. This way, the local
objective function abstracts the dependence on other nodes, eliminating the
need to consider their exact models or their raw
measurements. Nevertheless, the convergence of the local filter is
dependent on the rest of the network and we provide conditions that render
the $H_\infty$ convergence of the network of the filters. Furthermore, by
asserting further conditions on the individual performances of the other
nodes a guaranteed $H_\infty$ performance of the individual filters is
established. These conditions 
express that if all the 
neighbours maintain a certain level of accuracy then the local
filter also guarantees a nominated 
$H_\infty$ 
performance. 
 
To establish the above relationship, here we analyze the distributed filter
network consisting of estimators solving an auxiliary optimal filtering
problem at every node. In that sense, our approach bears some resemblance
with decentralized control where each controller is constructed to regulate
a local subsystem. The mentioned auxiliary filtering problem originates
in~\cite{Mortensen-1968,H80}; 
it was shown
in~\cite{ZU1a} to yield interconnected consensus-type filters that exchange
information between the network nodes although the
parameters for each filter could be computed online in the decentralized
fashion. 

The new element of this paper compared with~\cite{ZU1a} is how the
neighboring information is interpreted by each node. In~\cite{ZU1a}, each
node was considered to be agnostic about the amount of energy in the error   
 between the true state of the plant and the neighbours' estimates of that
 state. In contrast, here we consider a model where each node
 perceives a relationship between the energy in the neighbours' error and the
 accuracy of its own filter. 
 We give a detailed discussion of this
 idea later in the paper; for now we only note that technically our model
 adds a constraint on the energy in the error inputs arising in the auxiliary 
 minimum energy problems. Such a constraint has the form of an Integral
 Quadratic Constraint previously used in robust decentralized control
 problems and filtering problems; e.g., see
 \cite{LUO1,MSP9}. However, unlike those problems, the parameters of the
 constraints used here play the role of tunable  
 parameters which are adjusted according to the desired local and global
 performance. They also serve as indicators of sensitivity of the
 individual filters to the neighbours' performance. 

The main result of this paper are sufficient conditions on the network
parameters that ensure $H_\infty$ performance 
of the network consisting of the proposed minimum energy filters. As mentioned,
not only global disturbance attenuation is guaranteed by these conditions,
but also certain local $H_\infty$ properties of the node filters are
established. We show that these conditions admit the form of a convex
semidefinite  program, which enables constructing a filter network yielding a
suboptimal disturbance attenuation.

\paragraph*{Notation}
$\mathbb{R}^n$ is the Euclidean space of vectors, $\|\cdot\|$ is the
Euclidean norm, and for any positive semidefinite matrix $X$, $X=X'\ge 0$,
$\Vert a\Vert_X\triangleq (a'Xa)^{1/2}$. 
For $0<T\le \infty$, $\mathcal{L}_2[0,T)$ denotes the Lebesgue space of
vector-valued signals square-integrable on $[0,T)$. $\mbox{diag}[X_1
,\ldots, X_N ]$ denotes the block diagonal matrix with $X_1 ,\ldots, X_N$ as its diagonal blocks, and $\otimes$ is the
Kronecker product of matrices. $\lambda_{\min}(Z)$ is the smallest
eigenvalue of a symmetric matrix $Z$. 

\section{Problem Formulation and Preliminaries}\label{pf}

\subsection{The plant and the distributed estimator}\label{DistrEstProb}

Consider a linear system  
\begin{equation}
\label{state}
 \dot{x} = Ax +Bw, \quad x(0)=x_0,
\end{equation}
where $x\in\mathbb{R}^{n}$ and $w\in\mathbb{R}^{m}$ are,
respectively, the state and the unknown modeling disturbance input; the
latter 
is assumed to be $\mathcal{L}_2$ integrable on $[0,\infty)$. The matrices
$A\in\mathbb{R}^{n\times n}$ and $B\in\mathbb{R}^{n\times m}$ are known,
however the initial state $x_0$ is unknown and is considered to be part of
the uncertainty about the system (\ref{state}). 

The main objective of the paper is to determine conditions under which 
the plant state $x(t)$ can be estimated by a 
network of filters each using its plant measurement 
\begin{equation}\label{measurement}
y_i = C_ix + D_iv_i,
\end{equation}
where $i=1,2,\ldots,N$ indicates the measurement taken at node $i$ of the
network. Each measurement $y_i\in\mathbb{R}^{p_i}$ is imperfect, it is 
subject to a measurement disturbance $v_i$ taking values in
$\mathbb{R}^{m_i}$ that also 
belongs to the space $\mathcal{L}_2[0,\infty)$ by assumption. The coefficients
of each measured output are matrices of the matching dimensions, $C_i\in\mathbb{R}^{p_i\times n}$,
$D_i\in\mathbb{R}^{p_i\times 
  m_i}$, with $E_i\triangleq D_iD_i'>0$.

In addition to its direct measurements of the plant, each node 
receives information from other nodes of the network, of the form
\begin{equation}\label{communication}
c_{ij} = W_{ij}\hat{x}_j + F_{ij}\epsilon_{ij},
\end{equation}
where $\hat{x}_j$ is the estimate of state $x$ at the neighbouring node
$j$. The signal $\epsilon_{ij}$ with values in $\mathbb{R}^{m_{ij}}$ represents 
the communication errors or uncertainty in the communication channel, 
$\epsilon_{ij}\in \mathcal{L}_2[0,\infty)$. We assume that
$G_{ij}\triangleq  F_{ij}F_{ij}'>0$. 

The network graph describing communications between the filtering nodes is
assumed to be directed, its node and edge sets are denoted 
$\mathbf{V}=\{1,\ldots,N\}$ and $\mathbf{E}\subseteq \mathbf{V}\times
\mathbf{V}$, respectively. The neighborhood of node $i$, i.e., the set of
nodes which send information to node $i$, is denoted by
$\mathbf{N}_i=\{j:(i,j)\in\mathbf{E}\}$ and 
its cardinality is denoted $l_i$. The Laplace matrix of the network graph
is denoted $\mathbf{L}$~\cite{CH-1991}.  

Following~\cite{Olfati2005,SS-2009,U6} and many other papers on distributed
estimation, we 
consider a class of consensus-based interconnected filters each processing
the direct measurements $y_i$ and neighbours' information $c_{ij}$ by means
of a Luenberger-type observer of the form 
\begin{eqnarray}
\dot{\hat{x}}_i = A\hat x_i + L_i(y_i-C_i\hat x_i)
+\sum_{j\in\mathbf{N}_i}K_{ij}(c_{ij}-W_{ij}\hat x_i), 
\label{filter_i}
\\
\hat{x}_i(0)=\xi_i. \nonumber
  \end{eqnarray}

The estimation problem in this paper is to determine 
coefficients $L_i$, 
$K_{ij}$ (which can be time-varying) 
that ensure convergence of the network to trajectories of the
plant, and also guarantee an acceptable attenuation of the detrimental
effects of disturbances on the estimation error. Formally, these
properties are formulated as follows. 
Given a positive semidefinite matrix $P\in \mathbb{R}^{nN \times nN} $ and a
collection of positive semidefinite matrices 
$\mathcal{X}_i\in\mathbb{R}^{n\times n}$, and
constants $\gamma^2$ and $\bar\gamma_i^2$, $i\in \mathbf{V}$,  we wish to  
determine a collection of filters of the form (\ref{filter_i}) that
guarantee the following properties:
\begin{enumerate}[{\bf P1.}]
\item
In the absence of disturbances $w$, $v_i$ and $\epsilon_{ij}$, 
$j\in\mathbf{N}_i$, $i=1,\ldots,N$
the
estimation error of the  filter $e_i(t)=\hat{x}_i(t)-x(t)$ converges to
zero asymptotically.
\item
In the presence of disturbances, the network of filters (\ref{filter_i})
attains the type of $H_\infty$ disturbance attenuation property
  \begin{eqnarray}\label{H_inf}
\int_0^\infty \|e\|_P^2 dt &\leq& \gamma^2 \Bigg( \sum_{i=1}^N \Vert x(0)-\xi_i
 \Vert^2_{\mathcal{X}_i} + N\Vert w \Vert_2^2 \nonumber \\
 && +\sum_{i=1}^N \Big(\Vert
   v_i\Vert_2^2+\sum_{j\in\mathbf{N}_i}\Vert\epsilon_{ij}\Vert_2^2\Big)\Bigg),
\end{eqnarray}
where 
$e=[(\hat x_1-x)',\ldots,(\hat x_N-x)']'$ and $\Vert . \Vert_2^2$ is the $\mathcal{L}_2$ norm.   
\item
Provided the neighbours of node $i$ contribute a sufficient effort (this
will be quantitatively defined later) to assist $i$, it is also guaranteed that
at that node
\begin{eqnarray}
\lefteqn{\int_0^\infty\|e_i\|^2ds
\le \bar\gamma_i^2 \Big[\beta_i+
\|x_0-\xi_i\|_{\mathcal{X}_i}^2} && \nonumber \\
&&+\int_0^\infty\big[\|w\|^2+\|v_i\|^2+\sum_{j\in \mathbf{N}_i}
\|\epsilon_{ij}\|^2\big]ds \Big]; 
\label{problem.IQC.xtau.4}
\end{eqnarray}
$\beta_i>0$ is a constant which will be determined later. 
\end{enumerate}

These properties 
formalize the desired attributes of a distributed
filter that we want to achieve. In particular, property \textbf{P2}
specifies the desired global disturbance attenuation performance across the
sensor network using a network of decoupled filter equations
\eqref{filter_i}. Note that decoupled equations governing the gains in
filters~\eqref{filter_i} will be provided later. Furthermore, property
\textbf{P3} articulates the desired local disturbance attenuation provided
that there is sufficient contribution from the neighbours. The sufficient
contribution condition is quantitatively defined later in the paper.    

We remark that properties \textbf{P1}, \textbf{P2} jointly generalize the
property of $H_\infty$ consensus introduced in~\cite{U6}; also
see~\cite{LaU1,U8}. For example, let
$P=(\mathbf{L}+\mathbf{L}_{\top})\otimes P_0$ where $P_0=P_0'\ge 0$, 
and $\mathbf{L}_{\top}$ is the Laplacian matrix of the graph obtained from
the network graph by reversing its edges. This choice of $P$ results in the
left hand side of \eqref{H_inf} being equal to the weighted $H_\infty$
disagreement cost between the nodes, 
$\int_0^\infty\sum_i\sum_{j\in\mathbf{N}_i}\Vert
\hat{x}_i-\hat{x}_j\Vert_{P_0}^2ds $~\cite{U6,LaU1,U7}. More generally,
letting $P=(\mathbf{L}+\mathbf{L}_{\top})\otimes
P_0+\mathrm{diag}[P_1~\ldots~P_N]$, $P_i=P_i'>0$, reduces \textbf{P1},
\textbf{P2} to the property of strong robust synchronization introduced
in~\cite{U8}. In addition, property \textbf{P3} describes $H_\infty$
attenuation properties of individual node filters. Including such property
into analysis constitutes the main difference between the problem posed
above and the previous work in the area of distributed estimation.  

\subsection{Representation of the neighboring information}\label{Error.Dyn} 

To make performance analysis of individual 
filters possible, let us introduce the mismatch between the
disturbance-free information 
contained in the signal $c_{ij}$ and the corresponding true version of this
information,
\begin{equation}\label{xhatj}
 \eta_{ij}= W_{ij}(\hat{x}_j-x)=-W_{ij}e_j\in\mathbb{R}^{p_{ij}},  \quad j\in \mathbf{N}_i.
 \end{equation} 
With these signals the information
received by sensor $i$ can be represented as  
\begin{equation}\label{communication.alt}
c_{ij} = W_{ij}x + \eta_{ij}+F_{ij}\epsilon_{ij},\quad j\in\mathbf{N}_i.
\end{equation}
Equation (\ref{communication.alt}) can be regarded as an additional measurement of the
plant affected by disturbances $\epsilon_{ij}$ and $\eta_{ij}$. 

Treating the signals
$\eta_{ij}$, \emph{for the purpose of filter derivation}, as
the disturbances additional to  $w$,
$v_i$ and $\epsilon_{ij}$ has an effect of
decoupling node $i$ from its neighbours. Indeed, consider the error
dynamics of the filter (\ref{filter_i}) at node $i$,
\begin{eqnarray}
\dot{e}_i &=& \left(A-L_iC_i-\sum_{j\in \mathbf{N}_i}K_{ij}W_{ij}\right)e_i - Bw \nonumber \\
&& + L_iD_i v_i 
+\sum_{j\in\mathbf{N}_i}K_{ij}(\eta_{ij}+F_{ij}\epsilon_{ij}), 
\label{error_i.general} \\
\eta_{ki}&=&W_{ki} e_i, \quad i\in \mathbf{N}_k.\nonumber
  \end{eqnarray}
In this system, the signals $\eta_{ij}$, $j\in
\mathbf{N}_i$, play the role of exogenous disturbances and each signal
$\eta_{ki}=W_{ki} e_i$  represents the output used by agent $k$ for whom
$i$ is the neighbour, i.e., $i\in \mathbf{N}_k$. This interpretation
allowed us to construct in~\cite{ZU1a} minimum energy filters of the form
(\ref{filter_i}) with the property that for any initial
condition  $x_0$, arbitrary $\mathcal{L}_2$-integrable disturbances $w$,
$v_i$, $\epsilon_{ij}$, $\eta_{ij}$ and an arbitrary $T>0$
\begin{eqnarray}\label{Hinf.property.i}
\int_0^T\|e_i\|_{R_i}^2dt
&\le& \gamma^2\bigg(
\Vert x_0-\xi_i \Vert_{\mathcal{X}_i}^ 2 + \int_0^T\Big[ \Vert w \Vert^2  
+\Vert v_i\Vert^2 \nonumber \\
&&+ \sum_{j\in\mathbf{N}_i}(\Vert\epsilon_{ij}\Vert^2
+ \Vert \eta_{ij} \Vert^2_{Z^{-1}_{ij}})\Big]dt\bigg).
\end{eqnarray}
Here, $\gamma^2$ and $R_i=R_i'>0$ are a positive constant and matrices
whose existence is determined by certain LMI conditions
in~\cite{ZU1a}. Also, $Z_{ij}=Z_{ij}'>0$ are \emph{given} 
matrices; in~\cite{ZU1a} they were associated with the confidence of node
$i$ about performance of node $j$.  

Condition (\ref{Hinf.property.i}) provides an  $H_\infty$ type bound on the
energy in the filter estimation errors at node $i$ expressed in terms of
the energy of the disturbances affecting that node, 
and is similar to (\ref{problem.IQC.xtau.4}) in property \textbf{P3}. The
important difference between (\ref{Hinf.property.i}) and
(\ref{problem.IQC.xtau.4}) is that the former condition includes the energy in
the signals $\eta_{ij}$ that depend on the neighbours' accuracy. Also,
according to (\ref{Hinf.property.i}), the \emph{same level of disturbance
  attenuation} $\gamma^2$ is stated for all nodes. Our goal
is to revisit the design of the filters (\ref{filter_i}) to obtain a possibly
sharper $H_\infty$ property for at least some of the local filters, and for
other filters, to provide a 
means for assessing their local performance and sensitivity to
the neighbours' errors. 

\begin{figure}[t]
  \psfrag{w}{$w$}
  \psfrag{+}{$+$}
  \psfrag{ei}{$e_i$}
  \psfrag{vi}{$v_i$} 
  \psfrag{vj}{$v_j,\epsilon_{kj}~j\neq i$}
  \psfrag{epsi}{\hspace{-0.5cm}$F_{ij}\epsilon_{ij}$}
  \psfrag{epsji}{$F_{ki}\epsilon_{ki}$}
  \psfrag{etaij}{$\eta_{ij}$}
  \psfrag{etaki}{$\eta_{ki}$}
  \psfrag{Other}{\hspace{-.4cm}Other}
  \psfrag{subsystems}{\hspace{-.4cm}subsystems}
  \centering
  \includegraphics[width=6cm]{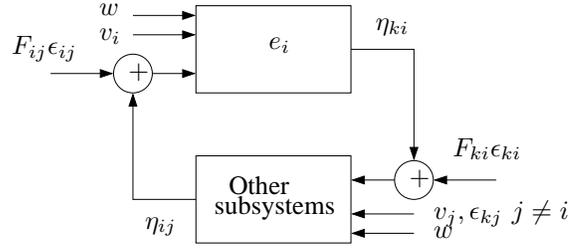}
  \caption{A two-block representation of the error dynamics system.}
  \label{two-blocks}
\end{figure}

Owing to the relation $\eta_{ij}=-W_{ij} e_j$, 
from the viewpoint of node $i$, the error dynamics
of the network can be seen as an interconnection of two systems,
representing, respectively, $i$'s own error dynamics and the 
errors dynamics of the rest of the system; see  
Figure~\ref{two-blocks}.
%
Motivated by (\ref{Hinf.property.i}), we propose the following
condition to formally capture the sensitivity of each node to the accuracy
of its neighbours' filters: 

\noindent \emph{For every $i$, there exist positive
definite symmetric matrices
$\bar Z_{ij}$ and constants $d_{ij}\ge 0$,
$j\in\mathbf{N}_i$, such that for all $t\in [0,\infty)$ and 
$w,v_i,\epsilon_{ij}\in\mathcal{L}_2[0,t)$,}  
\begin{eqnarray}
  \label{VU.IQC.i}
\lefteqn{\int_0^t\|\eta_{ij}\|^2_{\bar Z_{ij}^{-1}} dt  \le 
  \int_0^t ( \|e_i\|^2+\|w\|^2) dt + d_{ij} ,} \quad \\
&&\hspace{3cm}\forall j\in\mathbf{N}_i,i=1, \ldots,N.  \nonumber
\end{eqnarray}


As a generalized form of the property
(\ref{Hinf.property.i}), condition (\ref{VU.IQC.i}) reflects how the
neighbours' accuracy influences the local disturbance attenuation property
(\ref{problem.IQC.xtau.4}) at every node. Therefore in what follows, we
will use condition (\ref{VU.IQC.i}) to establish
(\ref{problem.IQC.xtau.4}), i.e., (\ref{VU.IQC.i}) is the quantitative
characteristic of the neighbours' effort mentioned in \textbf{P3}. 

To demonstrate  the role of (\ref{VU.IQC.i}) more vividly, take for example
$\bar Z_{ij}=\bar z_{ij} I$, with a scalar $\bar z_{ij}>0$ and suppose 
(\ref{problem.IQC.xtau.4}) holds provided (\ref{VU.IQC.i}) is satisfied with a 
very small $\bar z_{ij}$. Since according to (\ref{problem.IQC.xtau.4}),
the energy in $e_i$ is bounded, (\ref{VU.IQC.i}) suggests that node $i$ can only
tolerate relatively `small' mismatch inputs $\eta_{ij}$ to be able
guarantee (\ref{problem.IQC.xtau.4}). However,  a small energy in
$\eta_{ij}$ can only be accomplished by the corresponding neighbour $j$. 
This suggests that the eigenvalues of $\bar Z_{ij}$ may be indicative of
sensitivity of the local filters to fidelity of its neighbours'
estimates. In Section~\ref{tuning.IQCs} we will show that the matrices
$\bar{Z}_{ij}$  
can be computed jointly with the attenuation levels $\gamma^2$,
$\bar\gamma_i^2$. This provides the means for performance tuning of the
local filters. 

Similarly, the constants $d_{ij}$ in (\ref{VU.IQC.i})  
describe the bound on the estimation error energy that node $i$ is
prepared to tolerate from its neighbour $j$, in response to
(hypothetically) estimating the perfectly known plant ($w=0$) with the
utmost precision ($e_i=0$). Indeed, in this hypothetical case,
condition (\ref{VU.IQC.i}) reduces to a bound on the energy in the mismatch
disturbance signal $\eta_{ij}$ of the neighbour $j$.
  
\subsection{Distributed estimation problem}

We are now in a position to present a formal definition of the distributed
estimation problem described in Section~\ref{DistrEstProb}.  

\begin{problem}\label{Problem2}
Determine a collection of filters of the form (\ref{filter_i}) and matrices
$\bar Z_{ij}\in\mathbb{R}^{p_{ij}\times p_{ij}}$, $j\in\mathbf{N}_i$,
$i\in\mathbf{V}$, and 
constants $\gamma^2$ and $\bar\gamma_i^2$ such that the following
conditions hold: 
\begin{enumerate}[(i)]
\item
Given a positive semidefinite matrix $P\in \mathbb{R}^{nN\times nN}$, the
network of filters (\ref{filter_i}) achieves properties \textbf{P1} and \textbf{P2}
with this $P$ and the found $\gamma^2$.
\item
The following implication holds with the found $\bar
Z_{ij}$ and $\bar\gamma_i^2$:
If signals $\eta_{ij}(t)$, $j\in\mathbf{N}_i$, satisfy 
(\ref{VU.IQC.i}), then the filter (\ref{filter_i})
guarantees the satisfaction of condition (\ref{problem.IQC.xtau.4}), i.e.,
\textbf{P3} is satisfied.  
\end{enumerate}
\end{problem}

We stress that the global performance
properties \textbf{P1} and \textbf{P2} of the proposed distributed filter
will be proved without using condition (\ref{VU.IQC.i}). The IQC
(\ref{VU.IQC.i}) will only be 
used to guarantee certain local performance of each node $i$ subject to
acceptable performance of its neighbours. The latter development will be
analogous to how IQCs were used in the derivation
of decentralized robust controllers to quantify the uncertainty arising
from system interconnections; e.g, see~\cite{LUO1,PUSB}. However, different
from decentralized controllers in those references, our aim is to maintain
coupling between the filters, to ensure cooperation between them.

\section{Distributed minimum energy filtering with
  local performance guarantees}\label{RobustProb}  
 
In this section, our main results are presented. As was explained in
Section~\ref{Error.Dyn}, our goal is to obtain a converging (in the
$H_\infty$ sense) distributed filter which provides global estimation
performance described in item (i) of Problem~\ref{Problem2}, and also
characterize quantitatively the connection between local $H_\infty$
properties of the filters and their sensitivity to estimation accuracy of 
their neighbours. 


To solve Problem~\ref{Problem2}, we first introduce an auxiliary robust
minimum energy filtering problem involving a modified version of the 
standard minimum energy cost~\cite{Mortensen-1968}.
This cost functional, depending on the signals $w$ and
$\eta_i\triangleq 
 [\eta_{ij_1}'\ldots \eta_{ij_{l_i}}']'$, affecting the measurements
$y_i\vert_{[0,t]}$ and $\{c_{ij}\vert_{[0,t]}\;j\in\mathbf{N}_i\}$
available at node $i$ is as follows:
\begin{eqnarray}\label{costr.1}
\lefteqn{\bar J_{i,t}(x,w,\eta_i)} && \nonumber \\
&=& \frac{1}{2}\Vert x^{t,x}(0)-\xi_i \Vert_{\mathcal{X}_i}^ 2 + \frac{1}{2} \int_0^t\bigg( \Vert w \Vert^2  
+\Vert y_i - C_ix^{t,x}\Vert^2_{E_i^{-1}} \nonumber \\ 
&& +
\sum_{j\in\mathbf{N}_i}\Vert c_{ij} -
W_{ij}x^{t,x}-\eta_{ij}\Vert^2_{G_{ij}^{-1}} \nonumber \\
&&-\gamma^{-2}\Vert
x^{t,x}-\hat{x}_i\Vert^2_{R_i}\bigg)ds;
\end{eqnarray}

Compared to the standard minimum energy cost functional, it includes the
additional weighted penalty on the tracking error at node $i$; see the last
term in (\ref{costr.1}). It was shown 
in~\cite{ZU1a} that the inclusion of this term enforces a 
guaranteed $H_\infty$-type performance of the filter while a minimum energy
estimate is sought; cf.~\cite{McEneaney-1998}. The weight matrix of this
term, $R_i=R_i'>0$, $R_i\in\mathbb{R}^{n\times n}$ was regarded as
parameters of the filter, and a process of selecting those matrices to
optimize $\gamma^2$ was proposed in~\cite{ZU1a}. However,
different from~\cite{ZU1a}, the cost (\ref{costr.1}) does not include 
a direct quadratic penalty on $\eta_{ij}$. Instead, our derivation of the
local filters will impose the
constraint (\ref{VU.IQC.i}) on the mismatch signals $\eta_{ij}$, $j\in \mathbf{N}_i$. 

With these modifications, the auxiliary robust minimum-energy
filtering problem consists of determining a set of the
unknowns $x$, $w$, $\eta_i$ compatible with the measurements $y_i$ and the
communications $c_{ij}$ and minimizing the energy cost~\eqref{costr.1}
subject to the constraint (\ref{VU.IQC.i}):
\begin{equation}\label{problem.IQC}
\inf_{x} \left(\inf_{w\in \mathcal{L}_2[0,t]}\inf_{\eta_i\in \Xi_{i,t}} \bar
J_{i,t}(x,w,\eta_i)\right). 
\end{equation}
Here $\Xi_{i,t}$ denotes the class of vector signals $\eta_i$ obtained by
stacking up all $\eta_{ij}$, $j\in \mathbf{N}_i$, satisfying 
(\ref{VU.IQC.i}).
Originated from the minimum energy filtering~\cite{Mortensen-1968} and
least square fitting, this problem will lead to
the `most likely' minimum-energy trajectory $x^*_{i,t}(\cdot)$
compatible with the data at node $i$, $y|_{[0,t]}$,
$c_{ij}|_{[0,t]}$~\cite{Mortensen-1968}. The subscripts $i,t$ at
$x^*_{i,t}(\cdot)$ are to highlight that the trajectory $x^*_{i,t}(\cdot)$ is
consistent with the data collected on the interval $[0,t]$ at node $i$. By
definition, the end point of this trajectory is  the minimum-energy
estimate of the state $x(t)$, given the measurement data $y|_{[0,t]}$,
$c_{ij}|_{[0,t]}$: $\hat{x}_i(t)\triangleq x^*_{i,t}(t)$.

To solve the constrained optimization problem (\ref{problem.IQC}), we apply
the method of S-procedure~\cite{PUSB}.
In fact, since the cost $\bar J_{i,t}(x,w,\eta_i)$ itself depends on
$\hat{x}_i(\cdot)$, this requires us to solve a family of minimum energy
filtering problems, in which $\hat x_i$ is replaced with an arbitrary
signal $\bar x_i$. Then we take the fixed point of the mapping  $\bar
x_i(t)\to x^*_{i,t}(t)$ generated by this family of minimum energy
filtering problems,  as $\hat{x}_i(t)$. Due to lack of space, we omit
the details and proceed assuming that $\hat{x}_i(t)$ is such a
fixed point.
  
Let $\tau_i\in\mathbb{R}^n$ be a vector $\tau_i=[\tau_{i1}~\ldots
\tau_{iN}]'$ such that $\tau_{ij}> 0$ if
$j\in\mathbf{N}_i$, and $\tau_{ij}=0$ otherwise. Then define
\begin{eqnarray}\label{costr.tau}
\bar J_{i,t}^{\tau_i}(x,w,\eta_i) &=& \bar
J_{i,t}(x,w,\eta_i)+ \sum_{j\in\mathbf{N}_i}\frac{\tau_{ij}}{2} \int_0^t
  \Big(\|\eta_{ij}\|^2_{\bar Z_{ij}^{-1}} \nonumber \\
&-& \|x^{t,x}-\hat x_i\|^2-\|w\|^2 -\|y_i - C_ix^{t,x}\|^2\nonumber\\
&-&\sum_{r\in\mathbf{N}_i}\| c_{ir} -
W_{ir}x^{t,x}-\eta_{ir}\|^2
\Big)ds
\end{eqnarray}
and for fixed $t$ and $\hat x_i(\cdot)$, $x$, consider the unconstrained
optimization problem 
\begin{equation}\label{problem.IQC.tau}
\bar V_i^{\tau_i}(x,t)=\inf_{w,\eta_i\in \mathcal{L}_2[0,t]} \bar
J_{i,t}^{\tau_i}(x,w,\eta_i). 
\end{equation}
For each $t$, $x$, the optimization problem (\ref{problem.IQC.tau}) is 
a standard optimal tracking problem with a fixed terminal condition
$x(t)=x$, which has a unique solution under the condition 
\begin{eqnarray}\label{addit.LMIs}
\sum_{j\in \mathbf{N}_i}\tau_{ij}<1.
\end{eqnarray}
%
We now establish a relationship between this problem and 
the constrained inner optimization problem in~(\ref{problem.IQC}).

Let 
$
\mathcal{T}_i(t,x)\triangleq \{\tau_i \colon \mbox{(\ref{addit.LMIs})
  holds and}~\bar V_i^{\tau_i}(x,t) > -\infty\}.
$ 
Also for convenience, define
a vector $d_i\in\mathbb{R}^n$ whose $j$th component is $d_{ij}$ if 
$j\in\mathbf{N}_i$ and is 0 otherwise. 

\begin{lemma}\label{minimax} 
For every $\hat x_i(\cdot)$, $x\in\mathbb{R}^n$, if the corresponding set
$\mathcal{T}_i(t,x)$ is nonempty, then the value of the inner
optimization problem in (\ref{problem.IQC}) is finite,
\begin{equation}\label{problem.IQC.xtau}
\inf_{w\in \mathcal{L}_2[0,t], \atop \eta_i\in \Xi_{i,t}} \bar
J_{i,t}(x,w,\eta_i) \ge \sup_{\tau_i \in \mathcal{T}_i(t,x)} \left(\bar
  V_i^{\tau_i}(x,t) -\frac{\tau_i'd_i}{2}\right).
\end{equation}
\end{lemma}



From Lemma~\ref{minimax}, a
lower bound on the value of the problem~(\ref{problem.IQC}) follows:
\begin{eqnarray}
\lefteqn{\inf_x \inf_{w\in \mathcal{L}_2[0,t], \atop \eta_i\in \Xi_{i,t}} \bar
J_{i,t}(x,w,\eta_i)} && \nonumber \\
&& \ge \inf_x \sup_{\tau_i \in \mathcal{T}_i(t,x)} \left(\bar V_i^{\tau_i}(x,t) -\frac{\tau_i'd_i}{2}\right). 
\label{problem.IQC.xtau.1}
\end{eqnarray}
We now consider the following optimization problem
\begin{equation}
  \label{supinf.x}
\inf_x \bar V_i^{\tau_i}(x,t)=\inf_x\inf_{w,\eta_i\in\mathcal{L}_2} \bar J_{i,t}^{\tau_i}
(x,w,\eta_i).  
\end{equation}
A solution to this problem involves the differential Riccati equation 
\begin{eqnarray}
\lefteqn{\dot{Q}_i^{\tau_i} = Q_i^{\tau_i}A'+AQ_i^{\tau_i} -Q_i^{\tau_i}\Big( C_i'E_i^{-1}C_i} && \nonumber \\ 
&& 
+ \sum_{j\in\mathbf{N}_i}W_{ij}' \bar U_{ij}^{-1} W_{ij}
 - \gamma^{-2}R_i-\bar W_i \Big) Q_i^{\tau_i} +S_i , 
\label{summ_Riccati.tau} \\
\lefteqn{Q_i^{\tau_i}(0) = \mathcal{X}^{-1}_i,} \nonumber
\end{eqnarray}
where 
$\bar W_i= (\sum\limits_{j\in \mathbf{N}_i}\tau_{ij})I_n$,
$\bar U_{ij}\triangleq G_{ij} + \tau_{ij}^{-1} \bar Z_{ij}$,
$S_i= \Big(1-\sum_{j\in \mathbf{N}_i}\tau_{ij}\Big)^{-1}BB'$.



\begin{lemma}\label{Prop1.tau}
Given fixed $\tau_i\in \mathcal{T}_i(t,x)$ and $T>0$.
Suppose the differential Riccati equation 
(\ref{summ_Riccati.tau})  
has a symmetric nonsingular solution $Q_i^{\tau_i}=Q_i^{\tau_i}(t)$ 
on the interval $[0,T]$.   
Then the following filter computes recursively the minimizer 
$\hat{x}_i^{\tau_i}(t)$ of the optimization problem 
(\ref{supinf.x}) on the interval $[0,T]$, 
 \begin{eqnarray} 
\dot{\hat{x}}_i^{\tau_i} &=& A\hat{x}_i^{\tau_i} + Q_i^{\tau_i} 
\bigg(C_i'E_i^{-1}(y_i -
C_i\hat{x}_i^{\tau_i}) \nonumber \\ 
&&+\sum_{j\in\mathbf{N}_i}
W_{ij}'\bar U_{ij}^{-1}(c_{ij}- W_{ij}\hat{x}_i^{\tau_i}) \bigg),
\label{summ_obs.tau} \\
\hat{x}_i^{\tau_i}(0)&=&\xi_i.\nonumber
\end{eqnarray}
The value of the optimization problem
(\ref{supinf.x}) is finite and for $\hat x_i=\hat{x}_i^{\tau_i}$ 
is given by
\[
\bar \rho_{i,t}^{\tau_i} \triangleq 
\frac{1}{2}\int_0^t\bigg[\|y_i-C_i\hat
x_i^{\tau_i}\|^2_{E_i^{-1}} 
+ \sum_{j\in\mathbf{N}_i}\|c_{ij}-W_{ij}\hat
x_i^{\tau_i}\|_{\bar U_{ij}^{-1}}^2 \bigg]ds.
\]
\end{lemma}


Let 
\[
\bar{\mathcal{T}}_i(T)\triangleq \left\{\begin{array}{ll}\tau_i\colon & 
\mbox{(\ref{addit.LMIs}) holds and the
  DRE~(\ref{summ_Riccati.tau})}\\  
&  
\mbox{has a bounded positive definite}\\
& \mbox{solution on $[0,T]$.}
\end{array}
\right\}.
\] 

\begin{lemma}\label{tau.sets}
For all $T>0$,
$\bar{\mathcal{T}}_i(T)\subseteq
\bigcap\limits_{t\in[0,T], ~x\in\mathbb{R}^n}\mathcal{T}_i(t,x).
$
\end{lemma}


The following theorem summarizes the above discussion.

\begin{theorem}\label{T1}
Given constants $\gamma^2$ and
$\gamma_i^2$ and matrices $\bar Z_{ij}$, $j\in\mathbf{N}_i$, suppose 
the set $\bar{\mathcal{T}}_i(+\infty)=\bigcap_{T>0} \bar{\mathcal{T}}_i(T)$ is
not empty. Then for any $\eta_{ij}$ for which condition (\ref{VU.IQC.i}) holds,
the filter (\ref{summ_obs.tau}) computes recursively the process 
$\hat{x}_i^{\tau_i}(t)$ which satisfies condition
(\ref{problem.IQC.xtau.4}) with $\beta_i=\tau_i'd_i$.
\end{theorem}

Compared with the distributed minimum energy filter
in~\cite{ZU1a}, we have now obtained a family of 
suboptimal minimum energy filters for each node 
parametrized by $\tau_i\in\bar{\mathcal{T}}_i(+\infty)$. To be
able to apply Theorem~\ref{T1}, it is necessary to have a method for
computing at least one such vector $\tau_i$ for every node
$i$. 
In the next section, we will present an algorithm that 
accomplishes this
task. In addition, this algorithm obtains the matrices $\bar Z_{ij}$
and constants $\gamma_i^2$ consistent with the found $\gamma^2$, thus
providing a complete solution to  Problem~\ref{Problem2}. 

\section{Design of a robust distributed estimator}\label{tuning.IQCs}

The algorithm to compute a solution to Problem~\ref{Problem2}
utilizes a collection of linear matrix inequalities (LMIs) including the
condition  (\ref{addit.LMIs}) and the following matrix inequalities:
\begin{eqnarray}
  && \hspace{-.4cm}
\left[\begin{array}{cc}
{\small \begin{array}{l}A'\bar Y_i+\bar Y_iA+\big(\bar
      \gamma_i^{-2}+\sum\limits_{j\in \mathbf{N}_i}\tau_{ij}\big) I\\
      ~-\bigg(C_i'E_i^{-1}C_i+\sum\limits_{j\in \mathbf{N}_i}
                                     W_{ij}'\Upsilon_{ij}W_{ij}\bigg)
\end{array}}
& \bar Y_iB \\[20pt]
  B'\bar Y_i & \big(\sum\limits_{j\in
    \mathbf{N}_i}\tau_{ij}-1\big)I\end{array}\right]<0, \nonumber \\[10pt] 
  &&\bar Y_i=\bar Y_i'>0, \quad 
 \Upsilon_{ij}=\Upsilon_{ij}'>0, \quad \Upsilon_{ij}< G_{ij}^{-1},
\quad  
\tau_{ij}>0, \nonumber\\
&& \qquad\qquad \qquad\qquad \qquad\qquad j\in \mathbf{N}_i,\quad i=1,\ldots, N,
\nonumber \\
&& \bar\Theta >0. \label{LMI.tau}
\end{eqnarray}
Here, the symmetric
matrix  $\bar\Theta$ is composed as follows. Its 
diagonal blocks $\bar\Theta_{ii}$ are defined as
\begin{eqnarray*}
&&\bar\Theta_{ii} ={\scriptsize\left[\begin{array}{cccc}
\bar\theta^{ii}_0 & \bar\theta^{ii}_1 & \ldots &
\bar\theta^{ii}_{l_i} \\
(\theta^{ii}_1)' &
G_{ij_1}^{-1}& \ldots & 0 \\ 
\vdots & \vdots & \ddots & \vdots \\
(\theta^{ii}_{l_i})'  & 0 & \ldots & 
G_{ij_{l_i}}^{-1}
\end{array}\right]}, \\
&&
\bar\theta^{ii}_0=\sum\limits_{j\in\mathbf{N}_i}W_{ij}'\Upsilon_{ij}W_{ij}+
\big(\bar\gamma_i^{-2} +\sum_{j\in \mathbf{N}_i}\tau_{ij}\big)I-\gamma^{-2}P_{ii},\\
&&
\bar\theta^{ii}_k=W_{ij_k}'\Upsilon_{ij_k},
\quad k=1,\ldots, l_i.
\end{eqnarray*}
Also, its off-diagonal blocks $\bar\Theta_{ij}$, $i,j=1,\ldots,N$, $j\neq
i$, are  
\begin{eqnarray*}
&&\bar\Theta_{ij}=\begin{cases}\left[\begin{array}{cc}
\Psi_{ij}-\gamma^{-2}P_{ij} &
    \mathbf{0}_{n\times M_j}\\ \mathbf{0}_{n\times M_i} &
    \mathbf{0}_{M_i\times M_j}\end{array}\right], & i<j,\\[.3cm]
\bar\Theta_{ji}', &i>j,
\end{cases} 
\end{eqnarray*}
where
\begin{eqnarray*}
\Psi_{ij}={\scriptsize\begin{cases}
  -W_{ij}'\Upsilon_{ij}W_{ij}-W_{ji}'\Upsilon_{ji}W_{ji},
& j\in\mathbf{N}_i,i\in\mathbf{N}_j; \\
  -W_{ij}'\Upsilon_{ij}W_{ij},
& j\in\mathbf{N}_i,i\not\in\mathbf{N}_j; \\
  -W_{ji}'\Upsilon_{ji}W_{ji},
& j\not\in\mathbf{N}_i,i\in\mathbf{N}_j; \\
0 & j\not\in\mathbf{N}_i,i\not\in\mathbf{N}_j.
\end{cases}}
\end{eqnarray*}

The LMIs (\ref{LMI.tau}), (\ref{addit.LMIs}) represent a linear constraint
on the variables $\bar Y_i=\bar Y_i'>0$,  
$\bar\gamma_i^{-2}$, $\Upsilon_{ij}$, $\tau_{ij}>0$ ($j\in
\mathbf{N}_i$, $i=1,\ldots,N$), and 
$\gamma^{-2}$. Since $\gamma^2$ represents the disturbance attenuation
level in the distributed filter, a suitable set of filter 
parameters is of interest which minimizes this variable. This can be
numerically achieved by solving the convex optimization problem
\begin{equation}\label{SDP}
\sup \gamma^{-2} \quad \mbox{ subject to (\ref{LMI.tau}), (\ref{addit.LMIs}).}
\end{equation}
Let ${\gamma^*}^2$ be the value of the supremum in (\ref{SDP}).  

\begin{theorem}\label{stabt.tau}
Let the pair $(A,B)$ be stabilizable.
Given a positive semidefinite weighting matrix $P=P'\in\mathbb{R}^{nN\times
  nN}$, suppose $\gamma^2>{\gamma^*}^2$, $\tau_{ij}$, $\Upsilon_{ij}$,
$\bar\gamma_i^{-2}I$ and $\bar Y_i$, $j\in\mathbf{N}_i$, 
$i=1,\ldots,N$, are a feasible collection of matrices and scalars 
that satisfy the constraints of the convex optimization problem
(\ref{SDP}). Then each Riccati  
equation~\eqref{summ_Riccati.tau} with $R_i=(\gamma/\bar\gamma_i)^2$ has a
positive definite bounded solution on $[0,\infty)$. Furthermore, the
corresponding  
filtering algorithm~\eqref{summ_obs.tau},~\eqref{summ_Riccati.tau}
verifies claims (i) and (ii) of
Problem~\ref{Problem2}.  
\end{theorem}

As Theorem~\ref{stabt.tau} shows, solving the SDP problem (\ref{SDP})
allows us to determine the suboptimal $\gamma^2$ as well as the
local disturbance attenuation levels $\bar\gamma_i^2$ that characterize
local performance of the node filters (see
(\ref{problem.IQC.xtau.4})) as well as the matrices $\bar Z_{ij}$ in condition
(\ref{VU.IQC.i}) consistent with that performance. Then sensitivity of
performance of the obtained local filters to the neighbours' accuracy can
be assessed using, e.g., the eigenvalues of $\bar Z_{ij}$, as explained in
Section~\ref{Error.Dyn}. This process is illustrated in the example
presented next.


\section{Illustrative Example}\label{Simulations}

In this section, a simulated network of five sensor nodes is considered
that are to estimate a three-dimensional plant. The plant's state matrix and
the input matrix are 
\begin{equation}\label{A} 
 A = {\scriptsize\left[\begin{array}{ccc}
      -3.2  &  10  &  0  \\
      1 & -1 &  1 \\
      0 & -14.87 &  0 \\
     \end{array}\right]}, \quad
B={\scriptsize\left[\begin{array}{c}
    0.4\\
    0.4\\
    0.4
     \end{array}\right]}.
\end{equation}
The matrix $A$ corresponds to one of the regimes of the
controlled Chua electronic circuit considered in~\cite{U7}.  

The network consists of five nodes, its connectivity is
described by the set of directed edges
$\mathbf{E}=\{(1,3),(2,3),(3,1),(3,2),(3,4),(4,3),(4,5),(5,4)\}$.  
The matrices $C_i$ were taken from~\cite{U7} to be $C_1=C_4=0.001\times
[3.1923~ -4.6597~ 1]$ and $C_2=C_3=C_5=[-0.8986~ 0.1312~ -1.9703]$. 
Note that none of the pairs $(A,C_i)$ are observable, with $(A,C_1)$ and
$A,C_4)$ being not detectable. Also  following~\cite{U7}, all
communication matrices are taken to be $W_{ij}=I_{3\times 3}$ if 
$(i,j)\in \mathbf{E}$. Also, we let 
$D_i = 0.025I_{1\times 3}$ and $F_{ij} = 0.5I_{3\times 3}$.
  
\begin{table}[t]
\begin{center}
\caption{Solutions to the problem (\ref{SDP})}\label{table1}
  \begin{tabular}{|c|c|c|c|c|}
\hline
     & \multicolumn{2}{c|}{Simulation 1: $\bar Z_{ij}>0$} & 
       \multicolumn{2}{c|}{Simulation 2: $\bar Z_{ij}>0.1I$} \\
\hline
     & \multicolumn{2}{c|}{$\gamma^2=0.2500$} & 
     \multicolumn{2}{c|}{$\gamma^2=0.3116$} \\
\hline
Node & $\bar\gamma_i^2$ & $\min_j \lambda_{\min}(\bar Z_{ij})$ &
       $\bar\gamma_i^2$ & $\min_j \lambda_{\min}(\bar Z_{ij})$ \\
\hline
1 & 0.2643 & $2.6219\times 10^{-4}$ & 0.6288 & 0.1074 \\
2 & 0.0185 & 0.0250 & 0.0260 & 0.3416 \\
3 & 0.0181 & 0.0158 & 0.0395 & 0.1788 \\
4 & 0.1313 & $2.7548\times 10^{-4}$ & 0.2904 & 0.1000 \\
5 & 0.0176 & 0.0263 & 0.0265 & 0.2682\\
\hline
  \end{tabular}
\end{center}
\end{table}

For the above system two distributed filter designs were compared. Both
filters were designed to achieve a suboptimal 
$H_\infty$ consensus performance, that is, in (\ref{H_inf}) we selected
$P=(\mathbf{L}+\mathbf{L}_\top)\otimes I$, cf.~\cite{U6,LaU1}. First, the
optimization problem (\ref{SDP}) was solved with the above
parameters. Next, an additional constraint $\bar
Z_{ij}>0.1I$ was imposed. The computed levels of local $H_\infty$
attenuation $\bar\gamma_i^2$ and the minimum eigenvalues of the computed matrices
$\bar Z_{ij}$ with which the Property \textbf{P3} is guaranteed by
Theorem~\ref{stabt.tau} are shown in Table~\ref{table1}.  One can see that
in the first case, the filters at nodes 1 and 4 have much larger constants
$\bar\gamma_i^2$ and substantially smaller values of eigenvalues of
matrices $\bar Z_{ij}$. Together these features indicate that these filters are
significantly more sensitive to accuracy of their neighbours. This is not
unexpected given that the pairs $(A,C_1)$,  $(A,C_4)$ are not detectable. 
The second simulation indicates that robustness of the estimators
with respect to accuracy of their neighbours can be improved by
moderately increasing $\gamma^2$ and $\bar\gamma_i^2$.

 \section{CONCLUSIONS}\label{Conclusions}

 In this paper we proposed a distributed filtering algorithm by utilizing
 an $H_{\infty}$ minimum-energy filtering approach to the design of
 constituent filters. The algorithm employs
 a decoupled computation of the individual filter coefficients. This is
 achieved by considering the estimation error of neighbouring agents as
 additional exogenous disturbances weighted according to the nodes'
 confidence in their neighbours' estimates. The conditions are obtained
 under which the proposed filter to provides guaranteed internal 
 stability and desired disturbance attenuation of the network error
 dynamics. In addition each local filter  guarantees certain disturbance
 attenuation when assisted by the neighbours. We have also provided a
 simulation example that confirms convergence of the 
proposed filter in the case a system has undetectable pairs $(A,C_i)$ at
 some of the nodes. Tuning of the filter is discussed to reduce the
 dependence of the local filters from neighbours accurate estimates.


\newcommand{\noopsort}[1]{} \newcommand{\printfirst}[2]{#1}
  \newcommand{\singleletter}[1]{#1} \newcommand{\switchargs}[2]{#2#1}

\end{document}